\documentclass[twocolumn,twocolappendix]{aastex63}
\usepackage{amsmath}
\usepackage{abbrv}
\usepackage{hyperref}

\renewcommand{\log}{{\rm log_{\small 10}}}

\defcitealias{2020arXiv200616315K}{Paper~I}
\defcitealias{2018ApJ...853..173K}{KO18}

\graphicspath{{./}{paper-figures/}{paper-tables/}}

\shorttitle{Multiphase Galactic Wind Launching}
\shortauthors{Kim et al.}

\begin{document}

\title{A Framework for Multiphase Galactic Wind Launching using TIGRESS}

\author[0000-0003-2896-3725]{Chang-Goo Kim}
\affiliation{Department of Astrophysical Sciences, Princeton University, 4 Ivy Lane, Princeton, NJ 08544, USA}
\email{cgkim@astro.princeton.edu}
\affiliation{Center for Computational Astrophysics, Flatiron Institute, 162 Fifth Avenue, New York, NY 10010, USA}
\author[0000-0002-0509-9113]{Eve C. Ostriker}
\affiliation{Department of Astrophysical Sciences, Princeton University, 4 Ivy Lane, Princeton, NJ 08544, USA}

\author[0000-0003-3806-8548]{Drummond B. Fielding}
\affiliation{Center for Computational Astrophysics, Flatiron Institute, 162 Fifth Avenue, New York, NY 10010, USA}

\author[0000-0002-9849-877X]{Matthew C. Smith}
\affiliation{Harvard-Smithsonian Center for Astrophysics, 60 Garden Street, Cambridge, MA 02138, USA}
\affiliation{Center for Computational Astrophysics, Flatiron Institute, 162 Fifth Avenue, New York, NY 10010, USA}

\author[0000-0003-2630-9228]{Greg L. Bryan}
\affiliation{Department of Astronomy, Columbia University, 550 West 120th Street, New York, NY 10027, USA}
\affiliation{Center for Computational Astrophysics, Flatiron Institute, 162 Fifth Avenue, New York, NY 10010, USA}

\author[0000-0003-2835-8533]{Rachel S. Somerville}
\affiliation{Center for Computational Astrophysics, Flatiron Institute, 162 Fifth Avenue, New York, NY 10010, USA}
\affiliation{Department of Physics and Astronomy, Rutgers University, 136 Frelinghuysen Rd, Piscataway, NJ 08854, USA}

\author[0000-0002-1975-4449]{John C. Forbes}
\affiliation{Center for Computational Astrophysics, Flatiron Institute, 162 Fifth Avenue, New York, NY 10010, USA}

\author[0000-0002-3185-1540]{Shy Genel}
\affiliation{Center for Computational Astrophysics, Flatiron Institute, 162 Fifth Avenue, New York, NY 10010, USA}
\affiliation{Columbia Astrophysics Laboratory, Columbia University, 550 West 120th Street, New York, NY 10027, USA}

\author[0000-0001-6950-1629]{Lars Hernquist}
\affiliation{Harvard-Smithsonian Center for Astrophysics, 60 Garden Street, Cambridge, MA 02138, USA}

\begin{abstract}
Galactic outflows have density, temperature, and velocity variations at least as large as that of the multiphase, turbulent interstellar medium (ISM) from which they originate. 
We have conducted a suite of parsec-resolution numerical simulations using the TIGRESS framework, in which outflows emerge as a consequence of interaction between supernovae (SNe) and the star-forming ISM. 
The outflowing gas is characterized by two distinct thermal phases, cool ($T\lesssim10^4$~K) and hot ($T\gtrsim10^6$~K), with most mass carried by the cool phase and most energy and newly-injected metals carried by the hot phase. 
Both components have a broad distribution of outflow velocity, and especially for cool gas this implies a varying fraction of escaping material depending on the halo potential.  
Informed by the TIGRESS results, we develop straightforward analytic formulae for the joint probability density functions (PDFs) of mass, momentum, energy, and metal loading as distributions in outflow velocity and sound speed. 
The model PDFs have only two parameters, SFR surface density $\Ssfr$ and the metallicity of the ISM, and fully capture the behavior of the original TIGRESS simulation PDFs over $\Ssfr\in(10^{-4},1)\sfrunit$.
Employing PDFs from resolved simulations will enable galaxy formation subgrid model implementations with wind velocity and temperature (as well as total loading factors) that are based on theoretical predictions rather than empirical tuning.
This is a critical step to incorporate advances from TIGRESS and other high-resolution simulations in future cosmological hydrodynamics and semi-analytic galaxy formation models.
We release a python package to prototype our model and to ease its implementation.
\end{abstract}
\keywords{Galactic winds (572), Galaxy formation (595), Galaxy fountains (596), Galaxy winds (626), Stellar feedback (1602)}

\section{Introduction}\label{sec:intro}

Galactic scale outflows are prevalent in observations of star forming galaxies \citep[e.g.,][for reviews]{2005ARA&A..43..769V,2018Galax...6..138R} and play a central role in contemporary theory of galaxy formation and evolution \citep[e.g.,][for reviews]{2015ARA&A..53...51S,2017ARA&A..55...59N}. 
Although single-phase outflows have often been adopted in cosmological subgrid models \citep[e.g.,][]{2003MNRAS.339..289S,2006MNRAS.373.1265O,2013MNRAS.436.3031V}, real galactic outflows are clearly multiphase in nature \citep[][for a recent review]{2020A&ARv..28....2V}.  
Multiwavelength observations of fast-moving gas include radio lines from cold molecular and atomic outflows \citep[e.g.,][]{2015ApJ...814...83L,2018ApJ...856...61M}, optical and UV absorption lines from warm ionized outflows \citep[e.g.,][]{2000ApJS..129..493H,2005ApJ...621..227M,2015ApJ...809..147H,2016MNRAS.457.3133C,2017MNRAS.469.4831C}, and X-rays from hot ionized outflows \citep[e.g.,][]{1999ApJ...523..575L,2004ApJS..151..193S}. Furthermore, numerical simulations that resolve the multiphase ISM in galaxies and include supernova (SN) feedback \citep[e.g.,][]{2012MNRAS.421.3522H,2017MNRAS.466.1903G,2017ApJ...841..101L,2018ApJ...853..173K,2019MNRAS.483.3363H,2020ApJ...895...43S} show that both warm/cold and hot gas in the ISM are driven out together by superbubble expansion and breakout. Thus, launching of multiphase outflows appears to be the generic outcome of SN feedback in star-forming galaxies. 

In \citet[][\citetalias{2020arXiv200616315K} hereafter]{2020arXiv200616315K}, we analyzed the outflows in a suite of parsec-resolution numerical simulations spanning a range  of star-forming galaxy environments. 
We separated out two distinct thermal phases at $T<2\times10^4\Kel$ (cool) and $T>5\times10^5\Kel$ (hot) with a subdominant intermediate phase at temperatures in between. For each phase, we characterized  horizontally-integrated mass, momentum, energy, and metal fluxes and loading factors (fluxes normalized by the corresponding star formation rate (SFR), or by the SN momentum, energy, and metal injection rate).
We also measured horizontally averaged mean velocities of each outflow phase. In agreement with our  previous study  for solar neighborhood conditions \citep{2018ApJ...853..173K}, \citetalias{2020arXiv200616315K} showed that for all the environments investigated, (1) hot outflows deliver energy and SN-injected metals at high velocity to the circumgalactic medium (CGM); and (2) cool outflows carry much more mass, but at much lower velocity. We presented scaling relations for the dependence of multiphase outflow properties on the SFR, midplane pressure, and weight of the ISM, which are all (equally) good predictors for the mean outflow properties. 

The characterization of \citetalias{2020arXiv200616315K} addressed fundamental quantitative questions: how different are mass, momentum, energy, and metal outflow rates in different thermal phases? How do outflow rates scale with galactic conditions? 
However, in distilling ``velocity-integrated'' properties, important information regarding velocity and thermal distributions is lost. In particular, the cool-gas velocity distribution typically has an exponential wing extending to high velocity \citep{2018ApJ...853..173K,2020ApJ...894...12V}, such that significant cool ISM material could escape into the  CGM  region even if the mean cool outflow velocity is lower than a galaxy's escape speed.

Here we investigate the full joint probability distribution function (PDF) of outflow velocity and sound speed. We begin by showing that given a mass loading PDF, the momentum, energy, and metal loading PDFs can be constructed (\autoref{sec:PDFs}). We then develop a simple, parameterized model for the mass loading, with separate analytic functions describing cool and hot PDFs. These PDFs are combined with the scaling relations presented in \citetalias{2020arXiv200616315K} to create an easy-to-use outflow launching model (\autoref{sec:model}), which can be implemented in either semi-analytic or fully numerical cosmological models of galaxy formation. We provide a python 
package\footnote{\url{https://twind.readthedocs.io}; all figures in this \emph{Letter} are reproducible with the package.} \codename{} for model PDFs and sampling, and demonstrate its application (\autoref{sec:sample} and \autoref{sec:appendix_twind}).

It should be borne in mind that the particular set of TIGRESS models we employ have several advantages, but also come with caveats, as discussed below.  Thus, we consider the main goal of this \emph{Letter} to be a proof of principle: we show that joint PDFs of outflow velocity and sound speed are an efficient yet accurate way to encapsulate complex outflow properties from multi-physics, high-resolution simulations. We further show that an analytic model representation of the joint PDFs enables immediate and practical application of the results from small scale simulations to cosmological simulations and semi-analytic models.  
While the demonstration employs our current TIGRESS simulation suite, results from other simulations (with additional physics, and/or a wider parameter space) could be used in a similar fashion, fitting to obtain functional forms and parameters that characterize outflow PDFs based on kpc-scale galactic properties.

\section{Joint PDFs of outflow velocity and sound speed}\label{sec:PDFs}

We use a suite of local galactic disk models simulated with the TIGRESS framework \citep[][]{2017ApJ...846..133K}, as presented in \citetalias{2020arXiv200616315K}. The suite is comprised of 7 models, representing the range of galactic properties in nearby Milky Way-like star-forming galaxies, as summarized in \autoref{tbl:model}. The self-regulated disk properties cover a wide range of SFR surface density ($\Ssfr\sim 10^{-4}-1\sfrunit$), gas surface density ($\Sgas\sim 1-100\Surf$), and total midplane pressure/weight ($P_{\rm mid},\mathcal{W}\sim 10^3-10^6\,k_B\pcc\Kel$). We refer the reader to  \citetalias{2020arXiv200616315K} for full descriptions of models and methods (a brief summary can be found in \autoref{sec:appendix}).

We note here that stellar feedback processes considered in the TIGRESS framework include grain photoelectric heating by FUV radiation (without explicit radiation transfer) and SNe from star clusters and runaway OB stars, while other feedback processes including radiation pressure, photoionization, stellar winds, and cosmic rays are neglected. The missing feedback processes may affect the total outflow rates and distributions directly because some wind-driving forces such as cosmic ray and ionized-gas pressure gradients \citep[e.g.,][]{2018MNRAS.479.3042G,2018ApJ...865L..22E} are not represented, and/or indirectly, because early feedback might reduce clustering of star formation and SNe \citep[e.g.,][]{2020MNRAS.491.3702H}, which are known to enhance outflows \citep[e.g.,][]{2018MNRAS.481.3325F}. We note that the effects of ``early'' feedback has explicitly been  shown to be significant in dwarfs \citep{2018ApJ...865L..22E,2020arXiv200911309S}, but is not yet fully demonstrated in Milky Way-like conditions as simulated in the TIGRESS suite.
Also, the particular treatments in the TIGRESS framework for star formation using sink particles and SNe could potentially affect the properties of outflows (see \citet{2018ApJ...853..173K} and \citetalias{2020arXiv200616315K} for in-depth discussions). New metals (as opposed to the metals in the initial disk) in our simulations are injected only by SNe as we do not model stellar winds, which may affect the metallicity of outflows.

The above caveats and particularities certainly affect our specific quantitative results, but the overall approach we propose is quite general as a way to represent the mass, momentum, energy, and metals launched in multiphase outflows.

We now turn to distributions of outflowing gas 
in the TIGRESS suite.
Let $f_q(u,w;z)$ be the PDF of an outflow quantity $q$ at a given height $z$ within logarithmic velocity bins of vertical outgoing velocity $u\equiv\log v_{\rm out}$ and sound speed $w\equiv\log c_s$:
\begin{equation}\label{eq:pdf}
    f_q(u,w;z)\equiv \frac{1}{\abrackets{q(z)}}\frac{d^2 q(z)}{du dw}.
\end{equation}
Here, $f_q$ is in units of dex$^{-2}$,
$q(z)$ is a quantity defined at height $z$ over the entire $x$-$y$ horizontal domain and the time interval $t\in(t_1,t_2)$ of interest, and $\abrackets{q(z)}$ is the temporal and horizontal average of $q$ (also summed over all $u$ and $w$) so that the time averaged PDF has unit normalization, $\int f_q du dw =1$. The physical quantities $q$ of interest are vertical out-going fluxes
\begin{align}
    \mflux = \rho \vout 
    &\quad\textrm{(mass flux)},\label{eq:massflux}\\
    \pflux = \rho \vout^2 + P + \Pi_B
    &\quad\textrm{($z$-momentum flux)}, \label{eq:momflux}\\
    \eflux = \rho \vout \vB^2/2+\mathcal{S}_z
    &\quad\textrm{(energy flux)},\label{eq:eflux}\\
    \zflux = \rho Z \vout
    &\quad\textrm{(metal flux)}. \label{eq:Zflux}
\end{align}
Here,  $\vout\equiv v_z {\rm sgn}(z)$ is the vertical outgoing velocity,
\begin{equation}\label{eq:PiB}
    \Pi_B \equiv \frac{B^2}{8\pi} - \frac{B_z^2}{4\pi}
\end{equation}
is the vertical component of the Maxwell stress (magnetic pressure + tension),
\begin{equation}\label{eq:vB}
    \vB \equiv \rbrackets{v^2 + \frac{2\gamma}{\gamma-1}c_s^2}^{1/2}
\end{equation}
is the Bernoulli velocity, where we use isothermal sound speed $c_s^2\equiv P/\rho$, and 
\begin{equation}\label{eq:Sz}
    \mathcal{S}_z \equiv \frac{\rbrackets{\vout B^2 - B_{\rm out} \vel\cdot\Bvec}}{4\pi}
\end{equation}
is the vertical component of the Poynting flux with the vertical outgoing magnetic field $B_{\rm out} = B_z{\rm sgn}(z)$, and $Z$ is metallicity as traced by passive scalars in the MHD simulations.
Note that we adopt $\gamma=5/3$ so that $2\gamma/(\gamma-1)=5$. In the outflow analysis of  \citetalias{2020arXiv200616315K}, we did not include magnetic terms in momentum and energy fluxes; \autoref{eq:momflux} and \autoref{eq:eflux} include them for completeness but here we show they may be neglected.

The procedure to calculate the joint PDFs is as follows: we (1) extract one-zone thick slices at a distance from the midplane $|z|$ for both upper and lower sides  (either fixed heights at $z=\pm 500\pc$ and $\pm1\kpc$ or time-varying heights at $z=\pm H$ and $\pm 2H$, where $H$ is the instantaneous gas scale height) over $0.5<t/\torb<1.5$, (2) sum up each quantity within square bins $du=dw=0.02$~dex, and (3) normalize each PDF with the total ``outflowing'' quantity ($\vout>0$) averaged over the time range of interest at a given height defined by
\begin{equation}\label{eq:mean}
    \abrackets{q(z_k)} \equiv \frac{\sum_{n,i,j} q(x_i,y_j,t_n;z_k)\Theta(\vout>0)}{N_x N_y N_t},
\end{equation}
where $\Theta(C)$ is the top-hat-like filter that returns 1 if the conditional argument is true or 0 otherwise, $N_x$ and $N_y$ are the numbers of grid zones in the horizontal directions, and $N_t$ is number of snapshots analyzed. In \citetalias{2020arXiv200616315K}, we use $\overline{q}(z;t)$ to denote the horizontally-integrated/averaged quantities that are outflowing $\vout>0$ (with a phase separation if needed) at a given time. Thus, $\abrackets{q}$ here is simply  the  time-average of the corresponding $\overline{q}$, which is presented in Table~3 of \citetalias{2020arXiv200616315K} and available online at \href{http://doi.org/10.5281/zenodo.3872049}{doi:10.5281/zenodo.3872049}.

In addition to the total metal flux, it is of interest to quantify how enriched the outflow is compared to the ISM. To derive the distribution of the outflow enrichment factor, we first define the average metallicity within each logarithmic velocity bin as
\begin{equation}
    Z(u,w;z) \equiv \frac{\abrackets{\zflux(z)}}{\abrackets{\mflux(z)}}\frac{f_{\zflux}(u,w;z)}{f_{\mflux}(u,w;z)}.
\end{equation}
so that the corresponding enrichment factor is 
\begin{equation}
\zeta(u,w;z)\equiv \frac{Z(u,w;z)}{Z_{\rm ISM}}.
\label{eq:yZ}
\end{equation}
The mean ISM metallicity $Z_{\rm ISM}$ is  obtained by taking the time average of the instantaneous ISM metallicity $\overline{Z}_{\rm ISM}(t)$ defined by the mean metallicity of the cool phase within $|z|<50\pc$ (see Section 4.2 of \citetalias{2020arXiv200616315K}).

\begin{figure}
    \centering
    \includegraphics[width=\columnwidth]{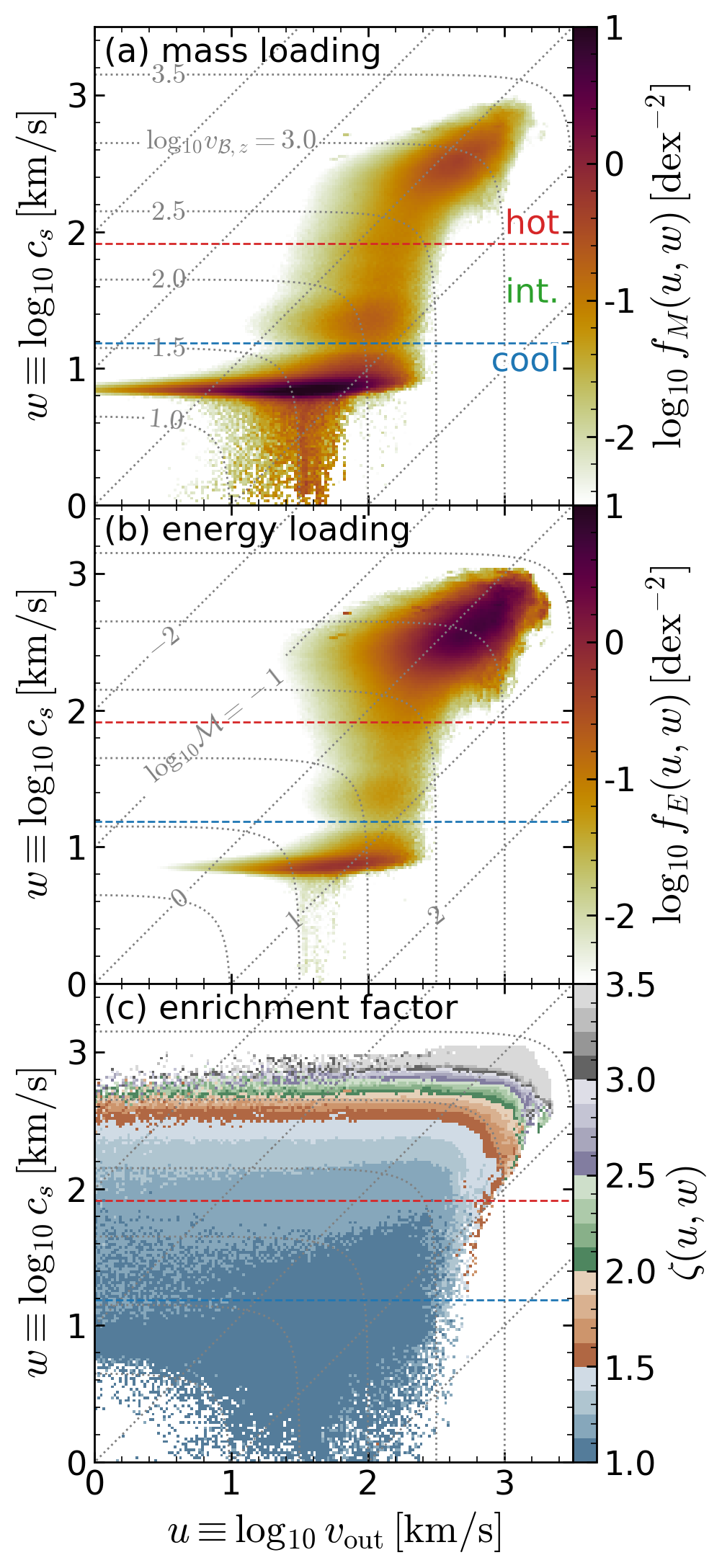}
    \caption{Joint PDFs in $\log \vout$ and $\log c_s$ for model R4 
    ($\Sgas\sim 30\Surf$ and $\Ssfr\sim 10^{-1}\sfrunit$)
    at $|z|=H$. 
    {\bf (a)} Mass loading PDF, {\bf (b)} energy loading PDF, and {\bf (c)} enrichment factor distribution. The red and blue dashed lines denote temperature cuts to separate cool ($T<2\times10^4\Kel$), intermediate ($2\times10^4\Kel<T<5\times10^5\Kel$), and hot ($5\times10^5\Kel<T$) phases. The dotted gray lines denote loci of constant $\vBz\equiv(\vout^2+5c_s^2)^{1/2}$ (labeled in (a)) and $\Mach \equiv \vout/c_s$ (labeled in (b));  loci are identical in all panels. 
    }
    \label{fig:R4-jointPDFs}
\end{figure}

To make our approach more general, we translate our results into loading factors, defined by ratios of outflow fluxes to corresponding areal rates of star formation (and related areal rates) in our simulations. This translation eases the connection to global simulations, in which e.g. the mass loading factor from a given cell represents the wind mass loss rate relative to SFR in that cell.  We note that $\Ssfr$ is still needed for our  model  parameterization; in global models appropriate projection and averaging may be used to define $\Ssfr$ in the region centered on a given cell (of arbitrary shape).

Following \citetalias{2020arXiv200616315K}, we define
\begin{equation}
    \eta_q (x,y,t;z) = \frac{\mathcal{F}_q(x,y,t;z)}{\abrackets{q_{\rm ref}\Ssfr/m_*}}.
\end{equation}
Here, $q=M$, $p$, $E$, and $Z$ as in \autoref{eq:massflux} -- \autoref{eq:Zflux}, $m_*$ is the mass of new stars formed per SN,  and the reference values per SN event for mass, momentum, energy, and metal mass are
\begin{align}
    \Mref = m_* &= 95.5\Msun, \label{eq:mref}\\
    \pref = E_{\rm SN}/(2v_{\rm cool}) &=1.25\times10^5\Msun\kms, \label{eq:pref}\\
    \Eref = E_{\rm SN} &=10^{51}\erg, \label{eq:Eref}\\
    \Zref\equiv M_{\rm ej}Z_{\rm SN} &= 2\Msun. \label{eq:Zref}
\end{align}
These values are adopted based on
a \citet{2001MNRAS.322..231K} initial mass function, with ejecta mass $M_{\rm ej}=10\Msun$, and metallicity $Z_{\rm SN}=0.2$ from STARBURST99 \citep[][]{1999ApJS..123....3L}. We choose $v_{\rm cool}=200\kms$ (see Section 4.1 of \citetalias{2020arXiv200616315K} for the full discussion of this choice and $\pref$).
With these definitions, $\eta_M$  is the ratio of mass outflow rate to SFR, $\eta_E$ ($\eta_Z$) is the ratio of energy outflow rate (metal mass outflow rate) to SN energy  (SN metal mass) injection rate, and $\eta_p$ is the ratio of  z-momentum outflow rate to the vertical momentum injection rate from post-Sedov-stage SNe.  
Note that the PDFs are identical for fluxes and loading factors as they differ by a constant factor and are normalized to be integrated to 1. Therefore $f_q$ with $q=M$, $p$, $E$, and $Z$ may denote either a flux PDF or loading PDF.
We note also that loading factors may be defined for all material in the outflow or separated by thermal phase, depending on whether the corresponding flux is for all material or phase-separated (see \citetalias{2020arXiv200616315K}). 

\autoref{fig:R4-jointPDFs} 
shows\footnote{An equivalent figure for all seven models and four values of $|z|$ can be created using \codename{}. The same is true for all other figures.}   
(a) the mass loading PDF ($f_{M}(u,w)$), (b) the energy loading PDF ($f_{E}(u,w)$), and (c) 
the metal enrichment factor 
($\zeta(u,w)$)
for model R4 at $|z|=H$.
As reported in \citetalias{2020arXiv200616315K} (see also \citealt{2018ApJ...853..173K}), it is evident that the cool outflow carries most of the mass (panel a), while the hot outflow carries most of the energy (panel b) and metals (panel c).
In addition, \autoref{fig:R4-jointPDFs}(b) clearly shows a wide distribution in $\vout$ (with a narrow spread in  $\cs$) for the cool outflow, contrasting in  \autoref{fig:R4-jointPDFs}(a) with a broader distribution along both axes for the hot outflow. This makes plain that naively adopting a single characteristic velocity and temperature would poorly represent both the mass and energy outflow rates.

For reference, \autoref{fig:R4-jointPDFs} includes contours of constant $\vBz$, where the outflowing component of the Bernoulli velocity is defined as
\begin{equation}
    \vBz\equiv(\vout^2+5c_s^2)^{1/2}.    
\end{equation}
To gauge whether fluid elements with given $(u,w)$ have sufficient energy to travel from the launching place to a distant location, $\vBz$ can be compared to the escape velocity $\vesc$ (which can be defined via the gravitational potential difference between wind launching position and the distant point).  

For outflows driven under conditions like model R4 (with $\Ssfr\sim0.1\sfrunit$), most hot outflows would escape the main galaxy if $\vesc\lesssim 300\kms$, delivering significant energy and metal fluxes far into the CGM. Cool outflows in a massive galaxy would, however, fall back as fountains. In the case of a low-mass galaxy with a shallow halo potential, e.g., $\vesc<50\kms$, cool outflows would carry significant mass outside the main galaxy into the CGM, at the same time as the hot wind carries energy and momentum.

\begin{figure*}
    \centering
    \includegraphics[width=\textwidth]{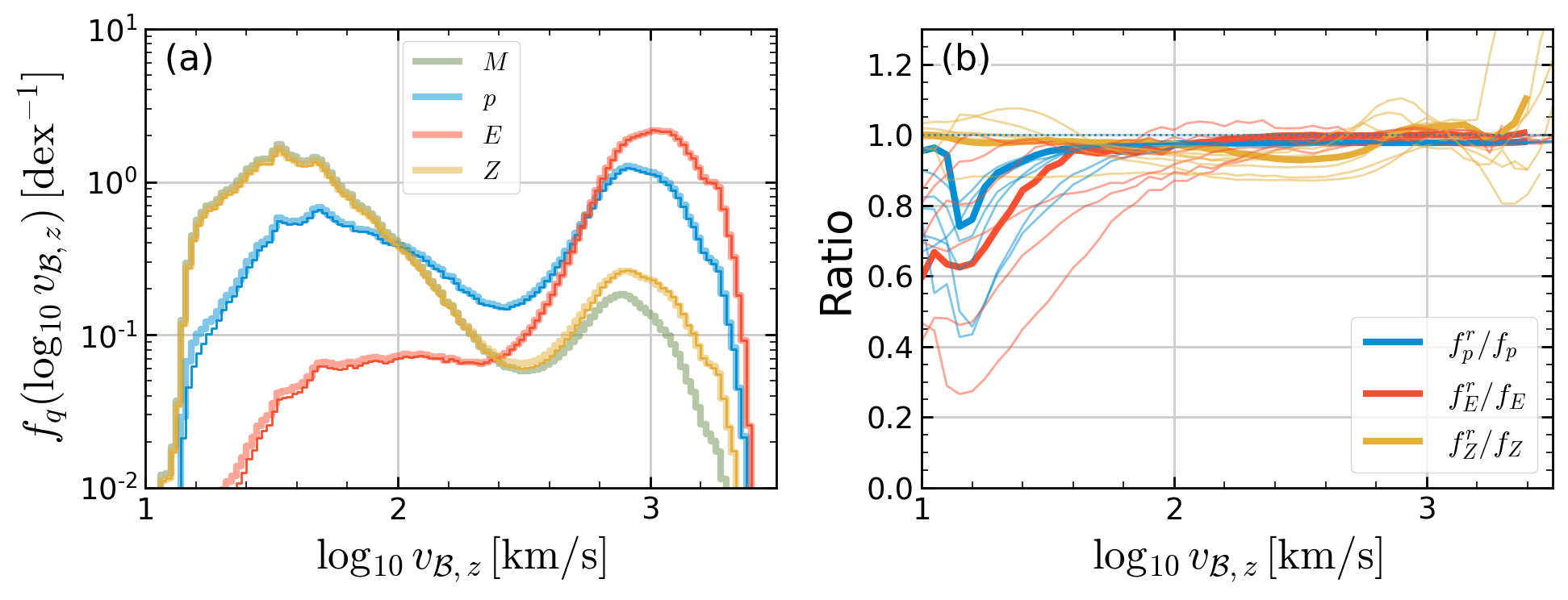}
    \caption{{\bf (a)} Examples of PDFs for 
    loading factors projected onto $\log \vBz$ for model R4 at $|z|=H$. 
    Thick lines show direct measurements of all PDFs, while 
    thin lines with the same color (overlying the thick lines almost everywhere) show reconstructions from the mass PDF of momentum (\autoref{eq:fpfromfM}), energy (\autoref{eq:fEfromfM}), and metal (\autoref{eq:fZfromfM}) PDFs. {\bf (b)} The ratios of reconstructed PDFs to the original PDFs for all models at $|z|=H$. 
    The mean ratio at a given $\log \vBz$ is obtained by a mass flux weighted average. (Thick lines correspond to model R4, shown to left.) 
    }
    \label{fig:recoveredpdf}
\end{figure*}

\autoref{fig:R4-jointPDFs}(c) demonstrates that the enrichment factor is tightly related to $\vBz$. This relation is rooted in the strong correlation between energy and SN-origin metal loading factors of the hot outflow seen in Figure 15 of \citetalias{2020arXiv200616315K} (see also \citealt{2015ApJ...800L...4C,2020ApJ...890L..30L}). 
We find that the model\footnote{Here and elsewhere we use a tilde to denote an analytic model, in which the parameters are determined by fits to the TIGRESS simulation suite outputs.} 
\begin{equation}\label{eq:yZmodel}
    \tilde{\zeta} = \rbrackets{\frac{\vBz}{3.2\times10^3\kms}}^{1.7}
    \sbrackets{\frac{Z_{\rm SN}}{Z_{\rm ISM}}-1}+1,
\end{equation}
for the enrichment factor defined in \autoref{eq:yZ} is in good agreement with the results for all heights ($|z|=H$, $2H$, 500~pc, and 1~kpc) where we measure the outflow properties from the full TIGRESS suite. As a result, for a given $Z_{\rm ISM}$, model outflow metallicity $\tilde{Z}\equiv \tilde{\zeta}Z_{\rm ISM}$ increases with the specific energy (or $\vBz$) at high-$\vBz$ and flattens to $Z_{\rm ISM}$ at low-$\vBz$. This formula predicts that, even in the limit of zero ISM metallicity (although this is outside the parameter space we explored), the hot outflow at high-$\vB$ would have large non-zero metallicity $\tilde{Z}\propto Z_{\rm SN}\vBz^{1.7}$ derived from very recent SN ejecta. Note that \autoref{eq:yZmodel} uses $\vBz$ which can be directly calculated from $u$ and $w$. \citetalias{2020arXiv200616315K} found a slight enrichment of the cool outflow ($\sim10\%$ at the largest $\Ssfr\sim 1\sfrunit$), but we neglect it for simplicity. In \autoref{eq:yZmodel}, $\tilde \zeta \rightarrow 1$ for low $\vBz$, so the outflow metallicity model is valid for cool gas, in which $\tilde{Z}$ approaches $Z_{\rm ISM}$. 

Under a certain set of assumptions, the momentum, energy, and metal loading PDFs can be recovered from the mass loading PDFs. If magnetic terms are negligible (and the bin size of PDFs is sufficiently small), the momentum loading PDF can be reconstructed from $f_{M}$ as 
\begin{equation}\label{eq:fpfromfM}
    f_p^r \equiv \frac{\abrackets{\eta_M}}{\abrackets{\eta_p}}
    \frac{\vout^2+c_s^2}{v_p\vout}f_{M},
\end{equation}
where $v_p\equiv \pref/\Mref = 1.3\times10^3\kms$.
The energy loading PDF can also be approximately reconstructed if the vertical component of kinetic energy dominates over the transverse component. In practice, there is non-negligible contribution from the transverse component of kinetic energy, but we find we can correct for this.  Our model results are consistent with a simple bias factor that describes the ratio of outflow component to total specific energy as a function of $\vBz$:
\begin{equation}\label{eq:bias}
    b \equiv \frac{\vBz^2}{\vB^2}= 0.1 \log \vBz + 0.6,
\end{equation}
for $\vBz$ in units of km/s and $\vBz\in (1,10^4)\kms$. 
We then obtain the reconstructed energy loading PDF from the mass loading PDF as
\begin{equation}\label{eq:fEfromfM}
    f_E^r \equiv \frac{\abrackets{\eta_M}}{\abrackets{\eta_E}}
    \frac{1}{2}\frac{\vBz^2}{v_E^2}\frac{f_{M}}{b},
\end{equation}
where $v_E\equiv (\Eref/\Mref)^{1/2} = 7.3\times10^2\kms$.
Similarly, the metal loading PDF can be recovered using \autoref{eq:yZmodel} as
\begin{equation}\label{eq:fZfromfM}
    f_Z^r \equiv \frac{\abrackets{\eta_M}}{\abrackets{\eta_Z}}\frac{Z_{\rm ISM}}{Z_e}\tilde{\zeta}(\vBz)f_{M},
\end{equation}
where $Z_e\equiv \Zref/\Mref\rightarrow 0.02$.

To demonstrate how well PDFs for other variables can be recovered from the mass loading PDF with Equations (\ref{eq:fpfromfM}), (\ref{eq:fEfromfM}), and (\ref{eq:fZfromfM}), \autoref{fig:recoveredpdf}(a) plots the original PDFs projected onto the  $\log\,\vBz$ axis (thick lines), in comparison with the reconstructed PDFs (thin lines), for model R4 at $|z|=H$. The reconstruction is successful: the thin lines are barely seen as they overlie the thick lines almost everywhere. For more quantitative comparison, \autoref{fig:recoveredpdf}(b) plots the ratios between reconstructed and original PDFs for all models at $|z|=H$ (thick lines are for R4 and thin lines for other models). 
Again, the recovery of all PDFs is quite good, especially at $\vBz$ larger than a few tens of  km/s (which is what matters in the outflow context). This justifies the general assumption that the magnetic stress is not important in outflows and confirms the validity of the enrichment factor model (\autoref{eq:yZmodel}) and the bias factor (\autoref{eq:bias}) for all cases.\footnote{We also confirm the same models can be applied to the results at all heights ($|z|=2H$, 500~pc, and 1~kpc).}

\section{Model PDFs and Validation}\label{sec:model}

\begin{figure*}
    \fig{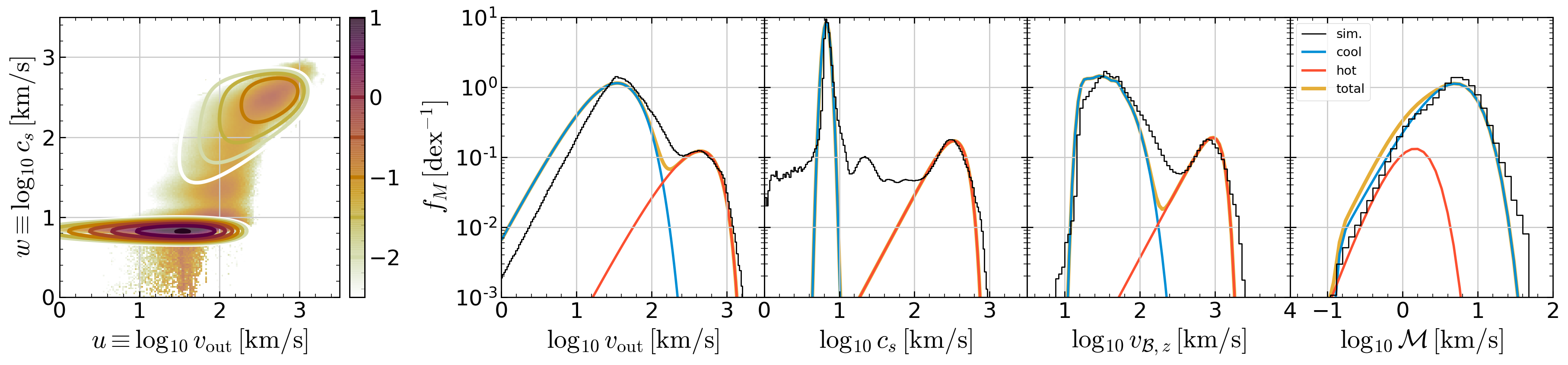}{\textwidth}{{\bf (a) Mass Loading}}
    \fig{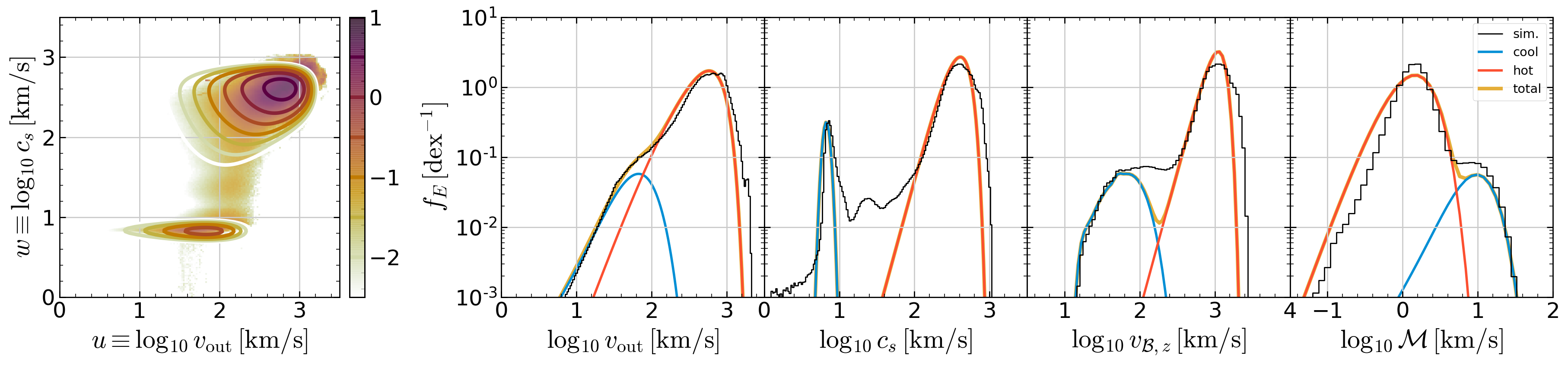}{\textwidth}{{\bf (b) Energy Loading}}
    \caption{Comparison between simulated and model PDFs for R4 
    at $|z|=H$:
    {\bf (a)} mass loading and {\bf (b)} energy loading. 
    In each row, the first column shows full joint PDFs in logarithmic color scale ($\log\, f_{M,E}\,[\rm dex^{-2}]$) from the simulation (color) and model (contour). The remaining four panels are histograms showing projections onto 
    (from left to right)
    $\vout$, $\cs$, $\vBz$, and $\Mach$ axes. Model PDFs are separated into cool (blue) and hot (orange) components. The sum of the two (yellow) matches simulated PDFs (black lines) well.}
    \label{fig:comp_model_R4}
\end{figure*}

\autoref{fig:R4-jointPDFs} and \autoref{fig:recoveredpdf} (see also \citetalias{2020arXiv200616315K}) make clear that the cool and hot gas have quite different loading properties, which suggests that for practical applications it will be  necessary to treat these components as two different \emph{species}.
To properly treat hot and cool winds on an individual basis, we first define an analytic two-component mass loading PDF model with an easy-to-use functional form that represents the results of the TIGRESS simulations well.\footnote{
For the purposes of quantifying winds, we do not separately analyze the coldest component ($T\lesssim 100$K), which may not be fully resolved in our simulations and does not contribute to mass, momentum, energy, and metal loading significantly.
}
We then combine this with the scaling relations for phase-separated loading factors (presented in Table 5 of \citetalias{2020arXiv200616315K}) and our reconstruction method (\autoref{eq:fpfromfM}, \autoref{eq:fEfromfM}, and \autoref{eq:fZfromfM}) to derive energy, momentum, and metal loading PDFs. We emphasize that the objective is not to describe every detail of the PDFs but to reasonably capture the overall behavior over the range of $\Ssfr/\sfrunit\in(10^{-4},1)$ covered in the simulation suite.  
With a goal of optimizing both physical fidelity and technical simplicity, we have found that just two free parameters are needed in our wind loading model: $\Ssfr$ and $Z_{\rm ISM}$. As we shall show, these two parameters encapsulate the essential aspects of local conditions of star-forming disks needed for characterizing wind properties.  

For the cool outflow ($T<2\times10^4\Kel$), we find that a model combining log-normal and generalized gamma functions,
\begin{equation}\label{eq:coolmodel}
\begin{split}
    \model^{\rm cool}(u,w) = &
    A_c \rbrackets{\frac{\vout}{\vmean}}^2
    \exp\sbrackets{-\rbrackets{\frac{\vout}{\vmean}}}\\
    &\exp\sbrackets{-\frac{1}{2}\rbrackets{\frac{\ln (c_s/\csmean)}{\sigma}}^2},
\end{split}
\end{equation}
describes the general shape of the distribution reasonably well.
Here, $A_c=(\ln 10)^2/(2\pi\sigma^2)^{1/2}=2.12/\sigma$. For this functional form, the mean outflow velocity is $2\vmean$. To fit the simulation PDFs 
at $|z|=H$,
we adopt constant values for $\csmean=6.7\kms$ and $\sigma=0.1$, while $\vmean$ is a function of $\Ssfr$:
\begin{equation}\label{eq:vmean}
    \frac{\vmean}{\kms} = 
    v_0 
    \Ssfrone^{0.23} + 3
\end{equation}
where 
$v_0=25$ and
$\Ssfrone\equiv\Ssfr/\sfrunit$.
The adopted form in \autoref{eq:vmean} differs slightly from the linear regression result presented in \citetalias{2020arXiv200616315K} (Eq. 57 there) to (1) adjust for a specific PDF shape adopted here, and (2) avoid arbitrarily low outflow velocity at very low $\Ssfr$. 
We find that only small adjustments are needed in the parameters to describe the simulation PDFs at larger $|z|$: $(v_0, \csmean/(\kms))=(45,7.5)$, $(45, 8.5)$, and $(60, 10)$ at $|z|=2H$, 500~pc, and 1~kpc, respectively.
This adjustment is physically reasonable because only the higher $\vBz$ components of cool outflows can travel farther.

For the hot outflow ($T>5\times10^5\Kel$), we construct a model mass PDF using two generalized gamma functions,
\begin{equation}\label{eq:hotmodel}
\begin{split}
    \model^{\rm hot}(u,w) = &
    A_h \rbrackets{\frac{\vBz}{\vBmean}}^2 
    \exp\sbrackets{-\rbrackets{\frac{\vBz}{\vBmean}}^4} \\
    &\rbrackets{\frac{\Mach}{\Machmean}}^3
    \exp\sbrackets{-\rbrackets{\frac{\Mach}{\Machmean}}}.
\end{split}
\end{equation}
The arguments  $\vBz=(\vout^2+5c_s^2)^{1/2}$ and  Mach number $\Mach\equiv\vout/c_s$ can be directly constructed from  $u$  and $w$.
Here, $A_h=(\ln 10)^2[2/\sqrt{\pi}]=5.98$.
For this functional form, the mean values of $\vBz$ and $\Mach$ are $0.69 \vBmean$ and $3\Machmean$.
We adopt a constant value of $\Machmean=0.5$, and a scaling relation for $\vBmean$: 
\begin{equation}\label{eq:vBmean}
    \frac{\vBmean}{10^3\kms} = 2.4\frac{\Ssfrone^{1/2}}{2+\Ssfrone^{1/2}} + 0.8.
\end{equation}
This adopted form is different from the linear regression presented in \citetalias{2020arXiv200616315K} (Eq. 60 there) to (1) keep $\vB\le v_{\rm ej}$ at high $\Ssfr$ and (2) accommodate a flattening at low $\Ssfr$.
The same hot outflow model works well at all four $|z|$ locations where the simulation PDFs are calculated. Given the generally high specific energy of the hot outflow, the shape of the PDF changes little within the range of $|z|$ we consider.
Note that \autoref{eq:coolmodel} and \autoref{eq:hotmodel} satisfy $\int \model(u,w) dudw=1$ individually. 

As models for the mass loading of the cool and hot phase 
at $|z|=H$,
we adopt power laws
\begin{align}
    \modelflux{M}{cool} = 0.85\,\Ssfrone^{-0.44}\label{eq:eta_M_cool}\\
    \modelflux{M}{hot} = 0.20\,\Ssfrone^{-0.07}\label{eq:eta_M_hot}
\end{align}
similar to the relations derived in \citetalias{2020arXiv200616315K} (see Fig. 8a,b there).  
As we are ignoring the intermediate component, the normalization in  $\modelflux{M}{hot}$ is slightly larger (by a factor 1.4) than that in \citetalias{2020arXiv200616315K}.  
We also note that  $\modelflux{M}{}\equiv\modelflux{M}{cool} + \modelflux{M}{hot}$ is not identical to the single combined power law fit shown in Fig. 13a of \citetalias{2020arXiv200616315K}. 
The model mass PDF obtained by combining the cool and hot components is given by
\begin{align}\label{eq:mfmodel}
    \model(u,w) = \frac{\modelflux{M}{cool}}{\modelflux{M}{}}\model^{\rm cool}
    +\frac{\modelflux{M}{hot}}{\modelflux{M}{}}\model^{\rm hot}
\end{align}

\autoref{fig:comp_model_R4} compares the simulated and model joint PDFs of model R4 in the $(u,w)$ plane (first column, color and contour, respectively) and projections along the $\log \vout$, $\log c_s$, $\log \vBz$, and $\log \Mach$ axes (from second to fifth column). Panels with projected PDFs show the combined model (yellow lines) and the individual cool (blue lines) and hot (red lines) components separately. The combined model PDF follows the original PDF from the  simulation (shown as black) reasonably well, modulo a dearth of intermediate temperature gas $w \in (1,2)$ and cold gas $w<0.5$ (but since these have low mass \emph{and} energy loading factors, this makes no practical difference).

We derive model PDFs for momentum, energy, and metal loading factors as
\begin{align}\label{eq:model}
    \tilde{f}_{q} = \frac{\modelflux{q}{cool}}{\modelflux{q}{}}\tilde{f}_{q}^{{\rm cool},r}
    +\frac{\modelflux{q}{hot}}{\modelflux{q}{}}\tilde{f}_{q}^{{\rm hot},r}
\end{align}
where $q=p$, $E$, and $Z$, and $\tilde{f}_{q}^{{\rm ph},r}$ is the reconstructed model PDF for each phase (ph=cool or hot) using Equations (\ref{eq:fpfromfM}), (\ref{eq:fEfromfM}), and (\ref{eq:fZfromfM}). 
As an example, for momentum 
\begin{equation}\label{eq:fpmodel}
    \tilde{f}_p^{{\rm ph}, r} = \frac{\tilde{\eta}_M^{{\rm ph}} }{\tilde{\eta}_p^{{\rm ph}}} 
    \frac{\vout^2+c_s^2}{v_p\vout}\tilde{f}_{M}^{\rm ph},
\end{equation}
with analogous expressions for $\tilde{f}_E^{{\rm ph},r}$ and $\tilde{f}_Z^{{\rm ph},r}$ based on Equations (\ref{eq:fEfromfM}) and (\ref{eq:fZfromfM}).
As we combine \autoref{eq:model} and \autoref{eq:fpmodel} (or analogous expressions for $q=E$ and $Z$), $\tilde{\eta}_{q}^{\rm ph}$ cancels out, and we only need models for the \emph{total} momentum, energy, and metal loading factors once we have constructed $\tilde{f}_{M}$ using the phase-separated $\modelflux{M}{cool}$ and $\modelflux{M}{hot}$.
We combine the power-laws in $\Ssfr$ for cool and hot phases 
at $|z|=H$
from Table 5 of \citetalias{2020arXiv200616315K}\footnote{
In order to construct model PDFs at different $|z|$, one should adjust the scaling relations of the loading factors to those from the given height, which are available at \href{http://doi.org/10.5281/zenodo.3872049}{doi:10.5281/zenodo.3872049}.
} to obtain total loading factors:
\begin{align}
    \modelflux{p}{} & = 0.04\Ssfrone^{-0.29}
    + 0.1\Ssfrone^{0.02}\label{eq:eta_p}\\
    \modelflux{E}{} & = 0.01\Ssfrone^{-0.12}
    + 0.2\Ssfrone^{0.14}\label{eq:eta_E}\\
    \modelflux{Z}{} & = 1.5\Ssfrone^{-0.36}
    + 0.42\Ssfrone^{0.04}.\label{eq:eta_Z}
\end{align}
We then renormalize the reconstructed model PDFs to make $\int \tilde{f}_q du dw =1$.

\autoref{fig:comp_model_R4}(b) shows the model energy loading PDF $\tilde{f}_E$ in comparison to the simulated energy loading PDF. Again, the agreement between the simulated and model PDFs are good.

\begin{figure*}
    \centering
    \includegraphics[width=\textwidth]{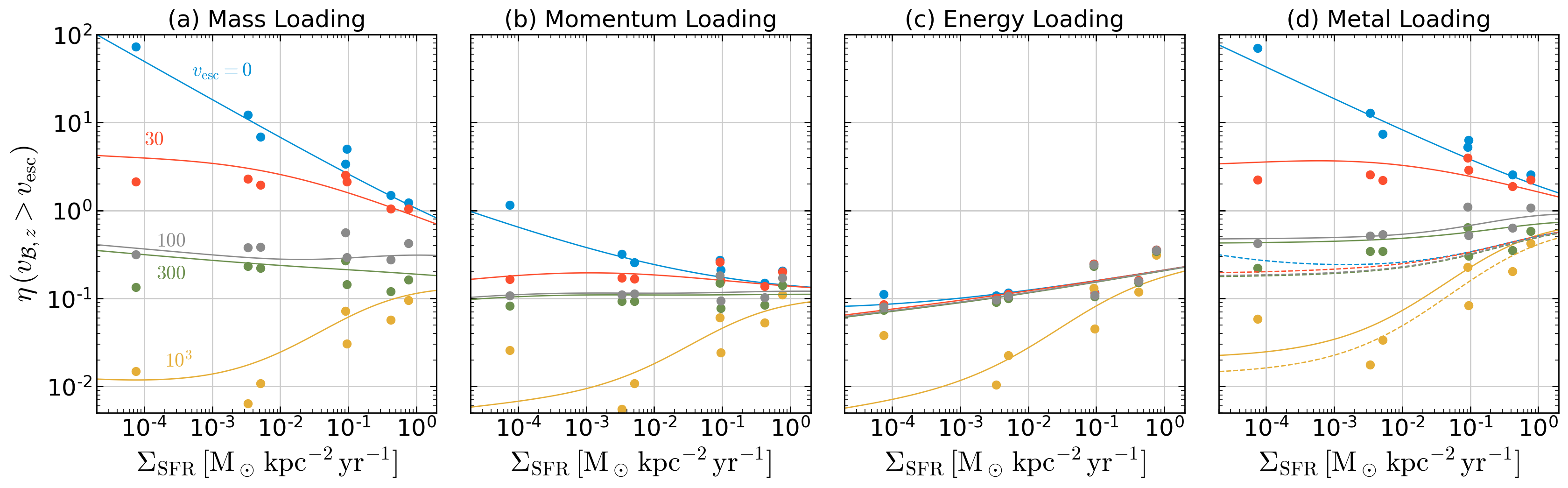}
    \caption{
    Loading factors for outflows with $\vBz>v_{\rm esc}$ (\autoref{eq:loading}) 
    at $|z|=H$. 
    Filled circles are directly calculated from the simulation PDFs, while solid lines are from the model PDFs. Solid and dashed lines in (d) denote the model loading factors for $Z_{\rm ISM}=0.02$ and 0, respectively.
    Overall, the model tracks the general behavior of the simulation results.
    }
    \label{fig:comp_vBcut}
\end{figure*}

To check the validity of the model PDFs over a range of $\Ssfr$, we compute loading factors for outflows for $\vBz>\vesc$, 
\begin{equation}\label{eq:loading}
    \eta_q(\vBz>\vesc)\equiv \tilde{\eta}_q
    \int_{\vBz=\vesc}^\infty \tilde{f}_{q}(u,w)dudw,
\end{equation}
where $q=M$, $p$, $E$, and $Z$. \autoref{fig:comp_vBcut} compares the model $\eta_q(\vBz>\vesc)$ at varying $\vesc=0$, 30, 100, 300, and $10^3\kms$ to direct results from the simulations 
at $|z|=H$.
These can be thought of as idealized outflow loading properties in halos with varying escape velocities. 
Direct results from the TIGRESS simulations are shown as filled circles, and the model compares well at all $\Ssfr$ and $\vesc$.

Note that for the purpose of this test, we normalize the model metal loading PDF for a fixed ISM metallicity, $Z_{\rm ISM}=0.02$. We plot as dashed lines in panel (d) the results for $Z_{\rm ISM}=0$, which is equivalent to the instantaneous SN-origin metal loading factors. This puts a floor  on the metal loading.

Despite its simplicity, the model correctly captures key behaviors from the simulations remarkably well. In particular, the high sensitivity of mass  loading to $\vesc$ (most extreme at low $\Ssfr$) and the general insensitivity of energy loading to $\vesc$ and $\Ssfr$ are notable.  
The former effect is due to the increase of the mass loading and the decrease of outflow velocities of cool gas at low $\Ssfr$, while the latter effect owes to the high outflow velocity and near-constant energy loading of hot gas produced by SNe.  
More subtle effects, such as the moderate decrease in energy loading at low $\Ssfr$  when $\vesc> 300\kms$, are also reproduced by the model.  We note that the energy-loading behavior of the model is mirrored in the metal-loading for $Z_{\rm ISM}=0$ because this is from SN ejecta, while the increase in metal loading at low $\vesc$  for $Z_{\rm ISM}=0.02$ is due to metal loss in low-velocity cool-ISM gas.

\section{Practical application}\label{sec:sample}

\begin{figure*}
    \centering
    \includegraphics[width=\textwidth]{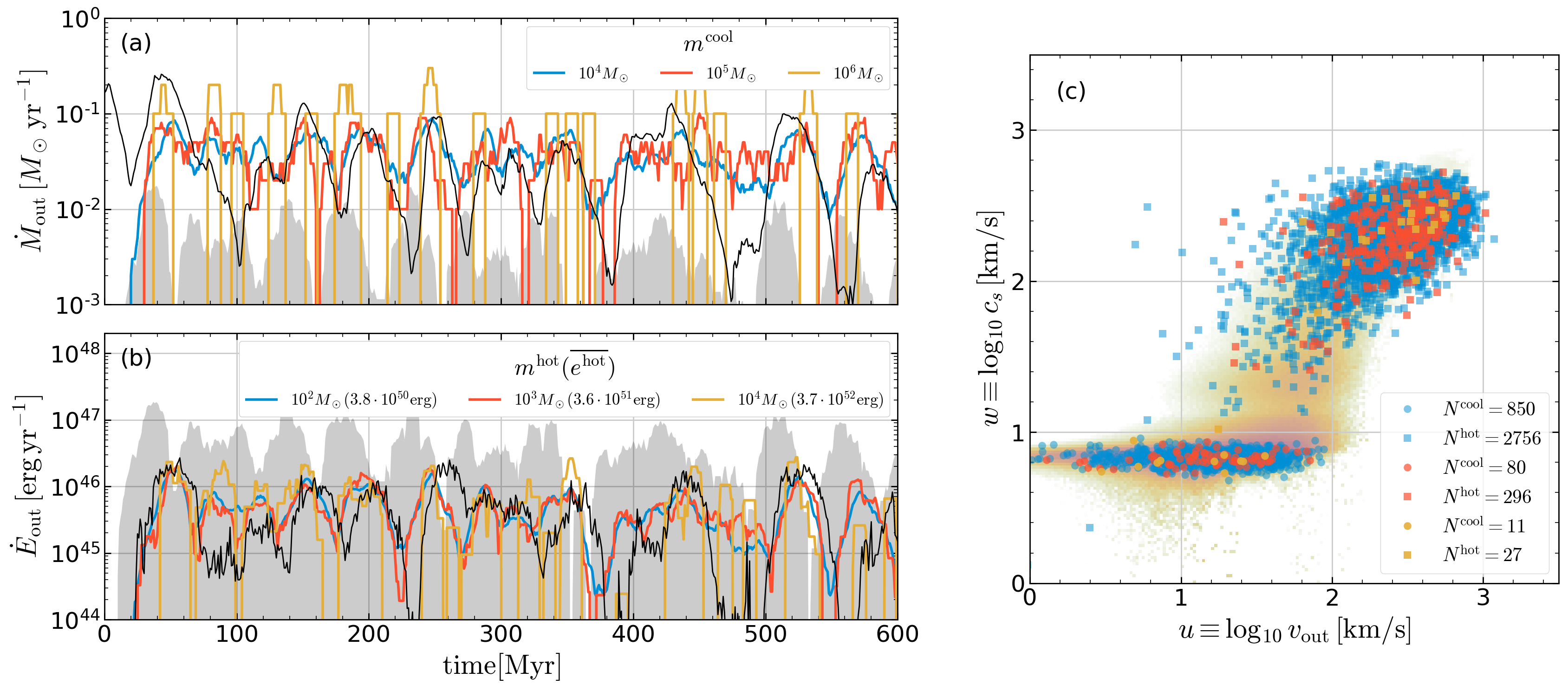}
    \caption{Model sampling demonstration for {\bf (a)} mass outflow rate of cool gas and {\bf (b)} energy outflow rate of hot gas 
    at $|z|=H$.
    The simulation result (black solid) is compared to the model for three different particle mass choices (colored lines; see keys).  
    The input to the model is $\Ssfr(t)$ from TIGRESS simulation R8
    (representing solar neighborhood conditions with $\Sgas\sim 10\Surf$ and $\Ssfr \sim 5\times10^{-3}\sfrunit$), 
    where $\SFR=\Ssfr L_x L_y$ is shown as the grey shaded region in (a) and the corresponding SN energy injection rate is shown as the grey region in (b).
    {\bf (c)} Distributions of cool (circles) and hot (squares) outflow particles sampled over $t =220$ - $440 \Myr$ from the different mass sampling cases (number of particles drawn is shown in the legend). The simulation PDF 
    at $|z|=H$
    over the same time interval is shown in the background.}
    \label{fig:test}
\end{figure*}

For practical implementation of our wind launching model, we release a python package \codename{} with our wind model. As cosmological simulations often launch winds as particles 
(e.g., \citealt{2006MNRAS.373.1265O,2013MNRAS.436.3031V}; but see \citealt{2008A&A...477...79D,2012MNRAS.426..140D,2014MNRAS.442.3013K} for alternative approaches),
the package also implements a particle sampling procedure. \codename{} is based on the two component PDF model at $|z|=H$ described in \autoref{sec:model}, but also supports models at different heights $|z|=2H$, 500~pc, and 1~kpc.

We demonstrate \codename{} following the procedure outlined in \autoref{sec:appendix_twind} using the R8 TIGRESS model, which has the longest simulation duration in our simulation suite. We treat our whole simulation box as equivalent to one resolution element in a cosmological simulation, giving a time series $\SFR = \Ssfr L_xL_y$ where $L_xL_y=(1024\pc)^2$ is the horizontal area of the simulation. For this demonstration, we fix $\Delta t=1\Myr$, and adopt constant mass quanta $m^{\rm cool}$ and $m^{\rm hot}$.  

\autoref{fig:test} shows (a) the mass outflow rate of cool  gas from the simulation (black) and the model with three different $m^{\rm cool}$; (b) the energy outflow rate of hot gas from the simulation (black) and the  model with three different $m^{\rm hot}$; and (c) the distribution of hot (squares) and cool (circles) particles for the different choices of particle masses. For simplicity, we only show mass (energy) outflow rates of the dominating cool (hot) component (see \autoref{fig:comp_model_R4}).

As the adopted SFR varies with time, the target mass outflow rate also fluctuates, with median  $3.7\times10^{-2}\Msun\yr^{-1}$ and  5(95)th percentile  $9.6\times10^{-3}$ ($6.4\times10^{-2})\Msun\yr^{-1}$. For our chosen $\Delta t=1 \Myr$, this translates into outflow mass $\sim 10^4-10^5\Msun$, mostly in cool gas. We thus expect complete, marginal, and incomplete sampling of cool outflows for $m^{\rm cool}=10^4\Msun$, $10^5\Msun$, and $10^6\Msun$, respectively. 
\autoref{fig:test}(a) is consistent with this expectation, with good temporal tracking at the lower two masses and poor tracking at the highest mass.\footnote{In both panels, model outflow rates are shifted by $10\Myr$ since there is a time delay between star formation/SN events near the midplane and gas outflows passing through one scale height above the midplane (see Appendix C in \citetalias{2020arXiv200616315K}).}  

For the R8 simulation, the 5 and 95th percentiles of the energy outflow rate are $4.9\times10^{44}$ and $1.4\times10^{46}\erg\yr^{-1}$, corresponding to a range $\sim 5\times 10^{50} - 10^{52}\erg$ for $\Delta t=1 \Myr$. Given the hot particle mass and model PDF, the mean particle energy is $\overline{e^{\rm hot}}\sim 4\times10^{50}$ for $m_{\rm hot}=10^2\Msun$ and increases nearly linearly with the particle mass. Therefore, we expect complete, marginal, and incomplete sampling of hot outflows for $m^{\rm hot}=10^2\Msun$, $10^3\Msun$, and $10^4\Msun$, respectively, as \autoref{fig:test}(b) demonstrates.

More generally, the necessary resolutions for a fair sampling of mass and energy outflow rates by cool and hot gas, respectively, are
\begin{align}
m^{\rm cool} &\sim \eta_M^{\rm cool}\dot{M}_*\Delta t\\
\frac{\overline{\vBz}^2}{2}m^{\rm hot} &\sim \eta_E^{\rm hot}E_{\rm SN} \frac{\dot{M}_*}{m_*}\Delta t,
\end{align}
where $\overline{\vBz}=0.69\vBmean$ is the mean Bernoulli velocity of the hot PDF (\autoref{eq:hotmodel}).
This implies that the ratio of resolution for hot and cool gas particles should obey
\begin{equation}
\frac{m^{\rm hot}}{m^{\rm cool}} \approx \frac{4 E_{\rm SN}}{m_* \vBmean^2} \frac{\eta_E^{\rm hot}}{\eta_M^{\rm cool}}
\end{equation}
which is $\sim 0.01 - 0.1$ in our simulations (increasing with $\Ssfr$). In simulations using particle-based codes, the hot wind particles would therefore generally need to be spawned with smaller mass than the common mass resolution of gas particles.

\section{Summary \& Outlook}

Outflows produced by SN feedback include both hot and cool phases, and even within a single phase there is a range of temperature and flow velocity.  Here, we extend the analysis of \citetalias{2020arXiv200616315K} using joint PDFs in  sound speed and outflow speed to characterize the mass, momentum, energy, and metal loading of the outflowing gas in a  TIGRESS simulation suite, separately treating cool and hot components.   
We demonstrate that the mass loading PDFs are well described by straightforward analytic expressions (\autoref{eq:coolmodel} and \autoref{eq:hotmodel}). The momentum, energy, and metal loading PDFs can then be reconstructed from the mass loading PDFs without significant loss of information. 
A sampling procedure utilizing our two-component PDF, prototyped in python as \codename{}, is able to successfully reproduce the  time-dependent simulated outflow rates in TIGRESS, provided that the respective mass and energy sampling of cool and hot phase outflows are sufficient to follow the true temporal evolution.

The framework developed in this \textit{Letter} for characterizing multiphase outflows using joint PDFs is quite general, and can be applied to any existing and future simulations in which multiphase outflows naturally emerge.
Additional feedback processes including cosmic rays, stellar winds, and radiation as well as additional physics (e.g., thermal conduction) or global geometry may alter the parameters compared to those calibrated using our existing  TIGRESS simulation suite. Different functional forms might be needed as well. Regardless of particular details, we consider the formalism we have introduced to analyze simulations and characterize joint PDFs of outflow velocity and sound speed as a fundamental advance in the representation of multiphase outflows.

The joint PDF model and sampling procedure outlined here can be applied to launch multiphase wind particles in cosmological simulations. This would require several changes with respect to current practices in big-box cosmological simulations. First, it is crucial to separately model hot and cool components, rather than a single component.  Second, the two components should have separate mass resolution, since resolving energy outflows in the hot gas requires a hot-gas particle mass 1 or 2 orders of magnitude lower than the particle mass required to resolve cool outflows.  

In particular, we consider sampling requirements for a solar neighborhood environment with $\Ssfr\sim3\times10^{-3}\sfrunit$, which is typical of star forming disks in both observations \citep[e.g.,][]{2020ApJ...892..148S} and simulations \citep[e.g.,][]{2020arXiv200616314M}. The mass resolution for baryons adopted in the Illustris-TNG50 simulations \citep{2019MNRAS.490.3196P,2019MNRAS.490.3234N} is $\sim 10^5\Msun$, which would be marginal for realizing mass outflow in cool gas but insufficient for realizing energy (and metal) delivery in hot outflows. 

The wind launching model outlined here requires just two parameters, the local ISM metallicity $Z_{\rm ISM}$ and the local star formation  rate per unit area  in the disk, $\Ssfr$.  The first is readily available in current cosmological simulation frameworks, but the latter typically is not. There are two different issues.  First, the disk scale height is generally not resolved in the current generation of large volume simulations.  As a result, the  true gas volume density (or pressure) is not known, and without knowledge of the corresponding internal dynamical timescales it is not possible to make a  physically-based prediction for the star formation rate on a cell-by-cell basis.  To address this issue, either the scale height must be resolved (e.g. in zoom simulations), or a sub-grid model for estimating the true gas  scale height must be included. Second, to obtain $\Ssfr$ on-the-fly for individual cells, additional computation involving some overhead (e.g., for neighbor searches) would be required. 

While the new approach to subgrid wind modeling we describe would involve technical challenges and computational costs, the return on the investment would be wind properties that represent local environments much more faithfully than current approaches. In particular, the usual practice in current cosmological simulations is to scale wind velocities relative to halo virial velocity \citep[e.g.,][]{2016MNRAS.462.3265D,2019MNRAS.486.2827D,2018MNRAS.473.4077P}, but this does not properly represent the physics of  cool gas acceleration, which mostly takes place at small scales within or near the disk in response to the \textit{local} rate of SN explosions. 
A wind launching model calibrated based on resolved local simulations would also be predictive and testable through, e.g., global correlations of galaxy properties, which is not the case for empirically-tuned subgrid models.

The results presented here are also of immediate practical use in semi-analytic models (SAMs) of galaxy formation \citep[e.g.,][]{2015MNRAS.453.4337S,2019MNRAS.487.3581F}. In contrast to traditional approaches adopted in SAMs, the inclusion of both mass and energy loading factors enables more sophisticated modeling. For example, many SAMs only account for the mass loss from the ISM due to outflows, and do not include the effects of energy deposited by winds. \citet{2020arXiv200616317P} have shown that preventative feedback due to energy deposition from stellar driven winds may be needed to allow SAMs to better reproduce the predictions from 
the FIRE-2 numerical hydrodynamic simulation suite \citep{2018MNRAS.480..800H}, 
especially in dwarf galaxies. Furthermore, a primary uncertainty in SAMs is what fraction of gas ejected by stellar-driven winds escapes the halo, and on what timescale this ejected gas returns to the halo. The $\vesc$ dependent loading factors presented here can be used to determine these quantities, thereby removing several of the free parameters that needed to be empirically calibrated in previous generations of SAMs.

Finally, we emphasize that our model only provides outflow properties at launching, close to the galactic disk ($|z|<1\kpc$). 
To understand and model the impact of multiphase outflows in the context of galaxy formation and evolution, it is necessary to follow \emph{wind interactions} with the CGM (which may be inflowing; e.g., \citealt{2019ApJ...872...47M,2020MNRAS.498.3664G}). 
These may or may not be resolved in cosmological simulations, or  explicitly modeled in SAMs. 
It is known from zoom-in simulations that there can be large differences between loading factors near the disk and after the interaction with the CGM  \citep[e.g.,][]{2015MNRAS.454.2691M,2017MNRAS.470.4698A,2019MNRAS.485.2511T}, which implies additional ``post-launch'' subgrid treatments would be required for lower-resolution large-box simulations.
Efforts are underway within the SMAUG collaboration to implement our \emph{wind launching} model, together with additional treatments for the interaction with the CGM,
both in numerical hydrodynamic simulations and in next generation SAMs.

\acknowledgements

This work was carried out as part of the SMAUG project. SMAUG gratefully acknowledges support from the Center for Computational Astrophysics at the Flatiron Institute, which is supported by the Simons Foundation. 
We appreciate the constructive report from the referee.
We are grateful to Ulrich Steinwandel, Viraj Pandya, Miao Li, and the rest of the SMAUG members for valuable discussions and helpful comments on the manuscript.
The work of C.-G.K. was supported in  part by a grant from the Simons Foundation (CCA 528307, E.C.O.). C.-G.K. and E.C.O. were supported in part by NASA ATP grant No. NNX17AG26G.
The work of M.C.S. was supported by a grant from the Simons Foundation (CCA 668771, L.E.H.)
G.L.B. acknowledges financial support from the NSF (grant AST-1615955, OAC-1835509).
Resources supporting this work were provided in part by the NASA High-End Computing (HEC) Program through the NASA Advanced Supercomputing (NAS) Division at Ames Research Center, in part by the Princeton Institute for Computational Science and Engineering (PICSciE) and the Office of Information Technology’s High Performance Computing Center, and in part by the National Energy Research Scientific Computing Center, which is supported by the Office of Science of the U.S. Department of Energy under Contract No. DE-AC02-05CH11231.

\software{{\tt Athena} \citep{2008ApJS..178..137S,2009NewA...14..139S},
{\tt astropy} \citep{2013A&A...558A..33A,2018AJ....156..123T}, 
{\tt scipy} \citep{2020SciPy-NMeth},
{\tt numpy} \citep{harris2020array}, 
{\tt IPython} \citep{Perez2007}, 
{\tt matplotlib} \citep{Hunter:2007},
{\tt xarray} \citep{hoyer2017xarray},
{\tt pandas} \citep{mckinney-proc-scipy-2010},
{\tt CMasher} \citep{CMasher},
{\tt adstex} (\url{https://github.com/yymao/adstex})
}

\appendix
\section{A brief summary of models and methods}\label{sec:appendix}
\restartappendixnumbering 

\begin{deluxetable}{lCCCCCC}
\tablecaption{Model Parameters\label{tbl:model}}
\tablehead{
\colhead{Model} &
\colhead{$\abrackets{\Sigma_{\rm gas}}$} &
\colhead{$\rho_{\rm sd}$} &
\colhead{$\torb$} &
\colhead{$R_0$} &
\colhead{$\Delta x$} &
\colhead{$\abrackets{\Sigma_{\rm SFR}}$}
}
\colnumbers
\startdata
R2    & 74  &      1 &  61 &  2 &   2 & 1.1\\
R4    & 30  &   0.45 & 110 &  4 &   2 & 1.3\times 10^{-1}\\
R8    & 11  &  0.092 & 220 &  8 &   4 & 5.1\times 10^{-3}\\
R16   & 2.5 &  0.005&  520 & 16 &   8 & 7.9\times 10^{-5}\\
LGR2  & 75  &   0.12 & 120 &  2 &   2 & 4.9\times 10^{-1}\\
LGR4  & 38  &  0.055 & 200 &  4 &   2 & 9.0\times 10^{-2}\\
LGR8  & 10  &  0.012 & 410 &  8 &   4 & 3.2\times 10^{-3}\\
\enddata
\tablecomments{
(1) model name. 
(2) gas surface density in $M_\odot\pc^{-2}$ averaged over $0.5<t/\torb<1.5$.
(3) volume density of stars and dark matter at the midplane in $M_\odot\pc^{-3}$. 
(4) orbit time $\torb\equiv2\pi/\Omega$ in Myr.
(5) galactocentric radius in kpc.
(6) spatial resolution of the simulation in pc. For all models, $(N_x,N_y,N_z)=(256,256,1792)$ grid zones are used.
(7) SFR surface density in $M_\odot\kpc^{-2}\yr^{-1}$ averaged over $0.5<t/\torb<1.5$.
}
\end{deluxetable}

The TIGRESS framework solves the ideal MHD equations on a uniform Cartesian grid using Athena \citep{2008ApJS..178..137S,2009NewA...14..139S}. We use a standard local shearing box in which $x=R-R_0$ and $y=R(\phi-\Omega t)$ are the local Cartesian coordinates at galactocentric distance $R_0$ (with box center rotating at angular speed  $\Omega$ about the galactic center), while $z$ is the global vertical coordinate centered at the disk midplane. The energy equation includes a net cooling term, $\mathcal{L}=n_H^2\Lambda(T)-n_H\Gamma$, where $n_H$ is the hydrogen number density,  $\Lambda(T)$ is the temperature dependent cooling coefficient at Solar metallicity adopted from \citet{2002ApJ...564L..97K} at $T<10^{4.2}\Kel$ and \citet{1993ApJS...88..253S} at $T>10^{4.2}\Kel$, and $\Gamma$ is the FUV-dependent heating rate due to photoelectric effect on small grains, allowing for plane-parallel attenuation of FUV radiation. The Poisson equation is solved to obtain the gravitational potential from gas and newly formed star clusters. Self-gravitating collapse is followed up to the density threshold ($>10^2 - 10^3\pcc$, higher at higher surface density models); above this density sink particles may be created to represent star clusters if flows are converging in all three directions and there is a local minimum of the gravitational potential. A simple stellar population synthesis model from STARBURST99 \citep{1999ApJS..123....3L} is used to obtain the FUV luminosity and SN rate of each cluster particle. SN events are modeled by either energy injection or momentum injection depending on the resolution and surrounding medium density \citep{2015ApJ...802...99K}. Our resolution is high enough to resolve the Sedov-Taylor stage of almost all SNe ($>90\%$), which is critical for multiphase outflow driving.

We note that our treatments of sink particles and SNe from them may enhance clustering of SNe, and the resulting ``burstiness'' may affect outflow properties. We convert all the gas in cells experiencing unresolved gravitational collapse into sink particles (assuming 100\% star formation efficiency within gravitationally collapsing cores). Each particle represents a star cluster with mass of $M_{\rm cl}$. Typical cluster masses are in the range of $10^3-10^5\Msun$, while a few $\sim 10^6\Msun$ clusters also form in the high gas surface density models, R2 and LGR2. Each cluster particle creates $N_{\rm SN}\sim M_{\rm cl}/m_*=10 (M_{\rm cl}/10^3\Msun)$ core collapse SNe over $\sim 40\Myr$. 2/3 of them explode at the position of the hosting cluster particle, while 1/3 are treated as OB runaways ejected from the hosting cluster at high velocity before explosion.

We use the standard model suite presented in \citetalias{2020arXiv200616315K}. The suite consists of 7 models, with key parameters summarized in \autoref{tbl:model}.

\section{{\tt Twind} Sampling Procedure}\label{sec:appendix_twind}

Given a total star formation rate $\SFR$, and wind mass quantum $m^{\rm ph}$ in each  phase ($\rm{ph}\in\{cool,hot\}$), the sampling procedure is as follows:
\begin{enumerate}
    \item Obtain the mass of the wind  for each phase:
    \begin{equation}
        M_{\rm wind}^{\rm ph} =  \modelflux{M}{ph}\SFR\Delta t,
    \end{equation}
    where $\Delta t$ is the time interval (which can be a simulation time step). 
    
    \item Draw an integer random variate $N^{\rm ph}$ from the Poisson distribution to choose the number of wind particles to spawn:
    \begin{equation}
        N^{\rm ph}\sim {\rm Pois}(k; \lambda),
    \end{equation}
    where ${\rm Pois}(k; \lambda) = \lambda^k e^{-\lambda}/k!$ is the Poisson distribution with the Poissonian mean $\lambda\equiv M_{\rm wind}^{\rm ph}/m^{\rm ph}$.
    
    \item Draw two random numbers $\xi_1$ and $\xi_2$ for each particle to assign $\vout$ and $c_s$. $\xi_1$ and $\xi_2$ are drawn either from a two-parameter generalized gaussian distribution (GGD; $G(x;d,p)$) and standard normal distribution ($g(x)$) for cool particles, or from two two-parameter GGDs for hot particles:
    \begin{align}
        \xi_1\sim G(x;2,1), \xi_2\sim g(x) \quad&\textrm{if ph=cool}\label{eq:cool_sample}\\
        \xi_1\sim G(x;2,4), \xi_2\sim G(x;3,1) \quad&\textrm{if ph=hot},\label{eq:hot_sample}
    \end{align}
    where $G(x;d,p) \equiv x^{d-1}e^{-x^p} p/\Gamma(d/p)$ is the two-parameter GGD and $g(x) \equiv  \exp(-x^2/2)/(2\pi)^{1/2}$ is the standard normal distribution. In practice, we tabulate the inverse of the CDF of three GGDs with parameters used in \autoref{eq:cool_sample} and \autoref{eq:hot_sample} and obtain $\xi_i=\cdf^{-1}(\xi_{U,i})$, where $\xi_{U,i}$ is a uniform random number in $[0,1)$.
    
    \item Assign $\vout$ and $c_s$:
    
    if ph=cool,
    \begin{equation}
        \vout=v_{\rm out,0}\xi_1, \quad c_s = c_{s,0}e^{\xi_2\sigma};
    \end{equation}
    and if ph=hot,
    \begin{align}
         c_s = \vBz/(\Mach^2+5)^{1/2}, \quad\vout = \Mach c_s,\\
        \textrm{where}\quad \vBz=\vBmean\xi_1, \quad\Mach = \Machmean \xi_2.
    \end{align}
    Here, $c_{s,0}$, $\sigma$, and $\Machmean$ are constants while $v_{\rm out,0}$ and $\vBmean$ are functions of $\Ssfr$ (see \autoref{sec:model}). To fit the simulation PDFs at different heights ($|z|=H$, $2H$, 500~pc, 1~kpc), we adjust these parameters and change $\tilde{\eta}_q$ from \citetalias{2020arXiv200616315K}.  These explicit parameterizations are implemented in \codename{}.
    
    \item Assign metals to the particle based on the metallicity in the launching region ($Z_{\rm ISM}$) and the metal enrichment factor (\autoref{eq:yZmodel})
    
    \item (Optional) Assign the velocity perpendicular to the outflow direction to the particle based on \autoref{eq:bias}:
    \begin{equation}
        v_{\perp}=\rbrackets{\frac{1-b}{b}}^{1/2}\vBz.
    \end{equation}
\end{enumerate}
In \autoref{sec:sample},
we draw samples with a fixed  particle mass as this is the typical practice in cosmological simulations, but it is also straightforward to draw samples for a fixed particle energy.

\bibliography{ref}
\end{document}